\documentstyle[12pt]{article}
\begin{document}

\newcommand{\la}[1]{\label{#1}}
\newcommand{\re}[1]{ (\ref{#1})}
\newcommand{\rf}[1]{ Fig. \ref{#1}}
\newcommand{\nn}{\nonumber}
\newcommand{\ed}{\end{document}}
\newcommand{\be}{\begin{equation}}
\newcommand{\ee}{\end{equation}}
\newcommand{\ba}{\begin{eqnarray}}
\newcommand{\ea}{\end{eqnarray}}
\newcommand{\baz}{\begin{eqnarray*}}
\newcommand{\eaz}{\end{eqnarray*}}
\newcommand{\bb}{\begin{thebibliography}}
\newcommand{\eb}{\end{thebibliography}}
\newcommand{\ct}[1]{${\cite{#1}}$}
\newcommand{\ctt}[2]{$^{\cite{#1}-\cite{#2}}$}
\newcommand{\bi}[1]{\bibitem{#1}}

\textheight=22.0cm
\hsize=15.8 cm
\topmargin=-1.05 cm
\oddsidemargin=0.5cm

\begin{titlepage}
\hspace*{10cm}{\bf FUB-HEP/93-99}

\vskip 3cm
\begin{center}
{\Large\bf Hard diffraction and proton spin problem}\\[1cm]
{N.I.Kochelev}\\[1cm]
{Institut f\"ur Theoretische Physik, Freie Universit\"at Berlin, \\
 Berlin, Germany
\footnote{On leave of absence from Joint Institute for Nuclear Research,\\
Laboratory of High Energy,
Head Post Office P.O.Box 79, SU-101000 Moscow, Russia\\
E-mail: kochelev@dec1.physik.fu-berlin.de }}
\end{center}
\begin{center}
{\bf Abstract}\\[0.2cm]
\end{center}

It is shown, that the hard component of quarks distribution
functions  induced by instantons
gives the essential contribution to hard diffractive processes. We predict
large polarization effects in these processes.
\vskip 5cm
\begin{center}
Submitted to: {\it Physics Letters B}
\end{center}
\end{titlepage}

In the conventional approach the diffraction is connected with pomeron
exchange Fig.1. In the framework of the perturbative QCD the "hard" pomeron
appears as result of the sum of a gluon ladder and it has the very high
intercept $\alpha_{hard}(0)\approx1.5$ \ct{a1}. This value is contradicted
 by the
experimental data on total, elastic and diffractive hadron-hadron
cross-sections which
support the conception of the "soft" pomeron \ct{a2} with the small
intercept $\alpha_{soft}(0)\approx1.08$. On the other hand, this small
intercept is contradicted by the new data on deep-inelastic structure
functions \ct{a3}, \ct{a4}. For the explanation of the anomalous growth of
these
structure functions
at low $x$ the hard pomeron is more suitable \ct{a5}. However, both of these
kinds of pomeron can not explain the date from NMC \ct{a6} and EMC \ct{a7},
where the large flavour and spin asymmetries of the sea quarks
distribution functions have been observed.

In paper \ct{a8} the new mechanism for anomalous behavior of the
structure functions has been proposed. It is related to the growth of the
quark-quark cross-section induced by nonperturbative fluctuation of
gluon fields-instantons \ct{a9}.

In this paper it was shown that the mixture in the nucleon wave function
 of  valence and sea quark components with high transfer momentum
 ($k_{\perp}\approx1GeV$),which is induced by instantons \ct{a10}, determines
 the growth of the structure function $F^N_2(x)$ at $0.0001<x<0.1$. At the same
 time  the strong dependence of this interaction \ct{a11}
\footnote{For simplicity, we present here only the instanton-induced
 lagrangian for case $N_f=2$. The case of $N_f=3$ was considered in ref.
 \ct{a12}.}
 \begin{eqnarray}
{\cal L}_{eff}^{(2)}(x)= \int d\rho\thinspace
 n(\rho)(\frac{4}{3}\pi^2\rho^3)^2\left\{\right.\relax
\bar u_Ru_L\bar d_Rd_L\left.\right. \relax
[1+\frac{3}{32}(1-\frac{3}{4}\sigma_{\mu\nu}^u\sigma_{\mu\nu}^d) \relax
\lambda_u^a\lambda_d^a]+(R\longleftrightarrow L)\left.\right\}
\la{e1}
\end{eqnarray}
from flavor and chirality, allows us to explain the observed large
violation of the Gottfried \ct{a13} and Ellis-Jaffe \ct{a14} sum rules.

Here, we will estimate the contribution of instantons to hard diffraction
\ct{a15} and to one-spin asymmetries in these processes.

The diagrams coming from the instanton-induced interaction to the
nucleon structure functions and diffraction are shown in Fig.2.
We estimate the instanton contribution by using the photon-pomeron
analogy \ct{a16} (Fig.2)
\be
\frac{sd^2\sigma^{inst}_{diff}}{\pi dtdM_x^2}=\frac{9\beta^4(F_1(t))^2}{4\pi^2}
(\frac{s}{M_x})^{2\alpha_P(t)-1}(1-\frac{M_x^2}{s})\bar F_2^{inst}(x),
\la{e2}
\ee
where $\beta^2=3.43 GeV^2$ is the  parameter which is determined from the
fit of elastic cross-sections, $F_1(t)$ is the Dirac formfactor of the proton,
 $x=Q^2/(M_x^2+Q^2)$, $Q^2=-t$ and
 \be
 \bar F_2^{inst}(x)=x(u_v^I(x)+d_v^I(x)+2\bar u^I(x)+2\bar d^I(x)+2\bar
s^I(x)).
 \la{e3}
 \ee
 In \re{e3} $q_i^I(x)$ is the instanton induced part of valence and
 sea quarks distribution functions.

In \ct{a8} from the consideration of the diagram Fig.2a the following
form for instanton induced valence and sea chirality and flavour
distribution functions were obtained:
 \ba
 2\bar u^I_+(x)&=&\frac{2}{3}\frac{N_Id_v(x)x^{\alpha_v}}{x^{\alpha_I}}f_u(x)
{\ },
 {\ }{\ }{\ }\hspace*{0.5cm}
 2\bar
u^I_-(x)=\frac{1}{3}\frac{N_Id_v(x)x^{\alpha_v}}{x^{\alpha_I}}f_u(x),\nn\\
 2\bar d^I_+(x)&=&\frac{1}{6}\frac{N_Iu_v(x)x^{\alpha_v}}{x^{\alpha_I}}f_u(x)
{\ },
 \hspace*{0.5cm}{\ }{\ }{\ }
 2\bar
d^I_-(x)=\frac{5}{6}\frac{N_Iu_v(x)x^{\alpha_v}}{x^{\alpha_I}}f_u(x),\nn\\
2\bar s_+^I(x)&=&2(\bar u_+^I(x)+\bar d_+^I(x)),
\hspace*{1.3cm}2\bar s_-^I(x)=2(\bar u_-^I(x)+\bar d_-^I(x)),\nn\\
  u^I_{v+}(x)&=&\frac{5}{6}\frac{N_Iu_v(x)x^{\alpha_v}}{x^{\alpha_I}}f_u(x){\
},
 {\ }{\ }{\ }\hspace*{0.5cm}
  u^I_{v-}(x)=\frac{1}{6}\frac{N_Iu_v(x)x^{\alpha_v}}{x^{\alpha_I}}f_u(x),\nn\\
  d^I_{v+}(x)&=&\frac{2}{3}\frac{N_Id_v(x)x^{\alpha_v}}{x^{\alpha_I}}f_u(x){\
},
 \hspace*{0.5cm}{\ }{\ }{\ }
 d^I_{v-}(x)=\frac{1}{3}\frac{N_Id_v(x)x^{\alpha_v}}{x^{\alpha_I}}f_u(x),
\la{e7}
\ea

 where $N_I$  is a constant,$ {\ }u_v(x),{\ }d_v(x)$ are the valence
 distribution functions :
 \be
u_v(x)=N_ux^{-\alpha_v}(1-x)^{bu},\hspace*{0.2cm}d_v(x)=
N_dx^{-\alpha_v}(1-x)^{bd},
\la{e5}
\ee
and
\ba
f_u(x)=\cases{ 1,&if $x> x_0$;\cr
 exp(-(x_0/x-1)),&$x \leq x_0$.\cr}
\label{e9}
\ea

The quark sea induced by the perturbative gluons does not differ in chirality
or  flavor, and therefore  it is  taken in the  form using
quark-counting rules:
\be
\bar u(x)_{+,-}^p=\bar d(x)_{+,-}^p=2\bar s(x)_{+,-}^p=
N_s(1-x)^5/x^{\alpha_{soft}(0)}.
\la{e11}
\ee

Chirality distribution functions for valence
quarks
have been chosen  in the form given by the Carlitz-Kaur model
\ct{a17}
\ba
\Delta u_v(x)&=&(u_v(x)-2d_v(x)/3)cos\Theta_D(x)\nn\\
\Delta d_v(x)&=&-d_v(x)cos\Theta_D(x)/3,
\la{e16}
\ea
where $ cos\Theta_D(x)$  takes into account the depolarization of
 the valence quarks in the low $ x $-region. The factor  $ cos\Theta_D(x)$
  can be interpreted as the contribution of confinement
   forces to the valence quark chirality violation.
It has the following form
\be
cos\Theta_D(x)=(1+H_1(1-x)^2/x^{\alpha_v})^{-1}.
\la{e17}
\ee

{}From \re{e7}-\re{e17} one can obtain the unpolarized
$\bar q(x)=\bar q_+(x)+\bar q_-(x) $  and the polarized
$\Delta\bar q(x)=\bar q_+(x)-\bar q_-(x) $
distribution functions.
The values of the parameters obtained by fitting of unpolarized and
polarized structure functions at $Q^2=5GeV^2$ are :
\ba
bu=2.79;{\ } bd=3.46;{\ }\alpha_v=0.37;{\ }N_s=0.095;{\ }\nn\\
N_I=0.0046;{\ }H_1=0.14;{\ }
x_0=0.0005;{\ }\alpha_I=1.36 .\nn
\ea

The region $-t=Q^2\geq1 GeV^2 $ is related to the hard diffraction region
\ct{a15} and
therefore the equation \re{e2} determines the contribution of instantons to the
hard diffractive cross-section.
The hard diffractive cross-section coming from the perturbative
 part of the proton
wave function is given by the same formula \re{e2}
 simply exchanging  $\bar F_2^{inst}(x)\rightarrow\bar F_2^{pert}(x)$,
  where $\bar F_2^{pert}(x)$ is the singlet
structure function without instanton contribution.
Thus the fraction of events induced by instantons in hard diffractive
processes is:
\be
\frac{d\sigma_{diff}^{inst}(x)}{d\sigma_{diff}^{tot}(x)}=\frac{\bar
F_2^{inst}(x)}
{\bar F_2^{pert}(x)+\bar F_2^{inst}(x)}
\la{e18}
\ee

In Fig.3 the result of the calculation of the instanton contribution
is shown. From this picture we conclude, that the large fraction
($\approx 30\%$) of hard diffractive
 processes is induced by instanton interaction.

The sea quarks distribution, induced by instantons, has the valence like
hard form at large $x$ \re{e7}. To our point of view, this hard
 component has been already observed in the UA8 experiment \ct{a18}.
In this experiment about $30\%$ of the observed hard diffraction events
can be explained only by the hypothesis of the existence of the very
hard sea component inside the proton.

 It should be mentioned also that the H1 and
ZEUS Collaborations have also observed the unusual events with
 the large rapidity gaps \ct{a4}. In the framework of the perturbative QCD
these diffractive events are high twist in the deep-inelastic
 scattering \ct{a19}.
Therefore they should have a very small cross-section. This prediction is in
 contradiction to experiment \ct{a4}, where  $5\div10\%$ events
have a large rapidity gap. In the framework of the nonperturbative QCD,
the instanton induced interaction gives the contribution to the DIS
diffraction and to the hadron-hadron diffraction in the same order
 (see Fig.2). It is connected with the structure of the lagrangian \re{e1},
 which includes the colourless exchanging part. This part can give the
 contribution to the DIS diffraction. The simple estimation of the matrix
 elements from the lagrangian \re{e1} shows that about $50\%$ from all events,
 induced by instantons, are without colour exchange between quarks. In
  the kinematical
 interval of the H1 and ZEUS Collaborations the instanton induced
 contribution to the $F_2^{ep}(x)$ structure function is  $10\div30\%$ Fig.4.
 Therefore, their contribution to the DIS diffraction is approximately
 $5\div15\%$ from
 all events with $10^{-1}<x<10^{-4}$. This value is in agreement with
 the preliminary results from the H1 and ZEUS Collaborations \ct{a4}.

It was pointed out above that interaction \re{e1} has a specific
dependence  from the chirality of the quarks. So, in the instanton field
the  chirality of the quarks is changed  by the value $\Delta Q=-2N_f$ .
The changing of the quarks chirality leads to an appearance of the quarks
orbital
 momentum. The direction of this orbital momentum is strong correlated with
 the direction of the spin of the proton \ct{a10}.
Thus the problem of how the transfer of the total angular momentum
 to the orbital momentum of the quarks has to be described, is solved and then
the proton spin problem has been decided.

The existence of the orbital  motion of quarks  leads to an appearance
 of one-spin
asymmetries in hadron-hadron interactions \ct{a20}, \ct{a21}.
In the framework of these models the values of these asymmetries are determined
by the
value of the reduction of the quark spin contribution to the proton spin.

Instantons give the essential contribution to the reduction of the quarks
spin contribution to the proton spin \ct{a8} and to hard diffractive processes.
 And
therefore in these processes  large polarized effects should be observed.

For example, we can estimate one-spin asymmetry in diffractive pp-collision
by using the following formula:
\be
A(x)\approx\frac{\Delta\bar F_2(x)}{\bar F_2(x)},
\la{e22}
\ee
where
\be
\Delta \bar F_2(x)=x(\Delta u(x)+\Delta d(x)+\Delta s(x))
\la{e23}
\ee
and $\Delta q(x)$ is the chirality carried by the proton quarks.

The result of the calculation of the one-spin asymmetry in hard
diffraction pp-processes is shown in Fig.5. At large $x$ the value of
the one-spin asymmetry is determined by the chirality carried by the
valence quarks and
at low $x$ region it is dependent from  the chirality of the sea quarks.
The value of the asymmetry at low $x$ is determined by
 the contribution of the instantons
and its large value comes from the large polarization of sea quarks in this
region.

It should be stressed that the isosinglet combination
\re{e23} is the fraction of the proton spin carried by the quarks. Therefore
 the polarized hard diffractive processes allow to measure this value.

It is also very important that in these processes we can measure the chirality
 distributions of quarks at very low values for $x$. For example, in the
experiment UA8 \ct{a18} the values of $M_x^2$ and $t$ were changing in
limits: $M_x^2=(2\div4)10^4GeV^2$; $-t=(0.9\div2.3)GeV^2$.  This corresponds
to the value $x=10^{-4}$! At present time such a small value of $x$ is not
accessible in DIS with polarized lepton beams .
 Therefore the measurement of one- and two-spin asymmetries in polarized
hard diffraction at RHIC Spin Collaboration \ct{a22} and at HERA \ct{a23}
can gives the unique possibility to find the decision of the proton spin
 problem.

The author is sincerely thankful to A.E.Dorokhov  for useful discussions,
 Prof.Meng Ta-chung for warm hospitality at
 the Free University of Berlin   and the DFG (Project: ME 470/7-1)
  for financial support.

\bb{99}
\bi{a1} L.N.Lipatov {\it Sov.\ Phys.\ JETP } {\bf 63} (1986) 904;\\
E.A.Kuraev, L.N.Lipatov, V.S.Fadin {\it Sov.\ Phys.\ JETP } {\bf 45} (1977) 199
\bi{a2} A.Donnachie, P.V.Landshoff {\it Nucl.\ Phys\ }{\bf B231} (1984) 189
\bi{a3} NMC {\it Phys.\ Lett.\ }{\bf B295} (1992) 159
\bi{a4} H1 Collaboration, Report on PANIC'93 Conference, Preprint DESY
93-113;\\
ZEUS Collaboration. Report on PANIC'93 Conference.
\bi{a5} A.D.Martin, W.J.Stirling, R.G.Roberts {\it Phys.\ Lett.\ }
{\bf B306} (1993) 145.
\bi{a6} NMC, P.Amaudruz et al.
 {\it Phys.\ Rev.\  Lett.\ }{\bf 66} (1991) 2712;\\
Preprint CERN-PPE/93-1993.
\bi{a7} EMC, J.Ashman et al.  {\it Nucl.\ Phys.\ }{\bf B328} (1990) 1.
\bi{a8} N.I.Kochelev Preprint FUB-HEP/93-13 (hep-ph/9307246), submitted to
{\it Phys.\  Lett.\ }.
\bi{a9} A.Ringwald {\it Nucl.\ Phys.\ } {\bf B330} (1990) 1;\\
O.Espinosa {\it Nucl.\ Phys. } {\bf B343} (1990) 310;\\
I.I.Balitsky, M.G.Ryskin {\it Phys.\ Lett.\ }{\bf B296} (1992) 185.
\bi{a10} A.E.Dorokhov,N.I.Kochelev
 {\it Phys.\ Lett.\ } {\bf B259} (1991) 335;\\
  {\it Phys.\  Lett.\ } {\bf B304} (1993) 167;
  {\it Int.\ J.\ of\ Mod.\ Phys.\ } {\bf A8} (1993) 603.
\bi{a11} 't Hooft {\it Phys. Rev.} {\bf D14} (1976) 3432.
\bi{a12} M.A.Shifman, A.I.Vainstein, V.I.Zakharov {\it Nucl.\ Phys.\ }
{\bf B163} (1980)~46.
\bi{a13} K.Gottfried {\it Phys.\ Rev.\ Lett.\ } {\bf 18} (1967) 1154
\bi{a14} J.Ellis, R.L.Jaffe, {\it Phys.\ Rev.\ }{\bf D9} (1974) 1444.
\bi{a15} H.Fritzsch, K.H.Streng {\it Phys.\ Lett.\ }{\bf B164} (1985) 391;\\
G.Ingelman, P.E.Schlein {\it Phys.\ Lett.}{\bf B152} 256;\\
E.L.Berger, J.C.Collins, D.E.Soper, G.Sterman
 {\it Nucl.\ Phys.\ }{\bf B286} (1987) 704;\\
A.Donnachie, P.V.Landshoff {\it Nucl.\ Phys.\ }{\bf B303} (1988) 634;\\
L.Frankfurt, M.Strikman {\it Phys.\ Rev.\ Lett.\ }{\bf 63} (1989) 1914;\\
J.C.Collins, L.Frankfurt, M.Strikman {\it Phys.\ Lett.\ }{\bf B307} (1993) 161.
\bi{a16}A.Donnachie, P.V.Landshoff {\it Nucl.\ Phys.\ }{\bf B244} (1984) 322
\bi{a17} R.D.Carlitz, J.Kaur {\it Phys.\ Rev.\ Lett.\ }{\bf 38} (1977) 673.
\bi{a18} UA8 Collaboration {\it Phys.\ Lett.\ }{\bf B297} (1992) 417.
\bi{a19} A.Donnachie, P.V.Landshoff {\it Phys.\ Lett.\ }{\bf B285} (1992)
172;\\
Hung Jung Lu, Joseph Milana {\it Phys.\ Lett.\ }{\bf B313} (1993) 234;\\
see also the last reference in [15].
\bi{a20} N.I.Kochelev, M.V.Tokarev {\it Phys.\ Lett.\ } {\bf 309} (1993) 416.
\bi{a21} C.Boros, Liang Zuo-Tang, Meng Ta-chung {\it Phys.\ Rev.\ Lett.\ }
{\bf 70} (1993) 1751.
\bi{a22} RHIC Spin Collaboration,G Bunce et al.
 {\it Part.\ World\ }{\bf 3} (1992) 1 .
\bi{a23} HERMES Collaboration, report on PANIC'93 Conference;\\
 Wolf-Dieter Nowak, Single spin asymmetries in proton-proton
 scattering at HERA energies, preprint DESY 93.
\eb
\newpage
{{\Large \bf Figure Captions:}}
\vskip 1cm
{{\large \bf Fig.1}}: Hard diffractive process.
\vskip 1cm
{{\large \bf Fig.2}}: The contribution to the a) DIS structure functions
and b) diffraction due to
 instantons .The circle is an instanton.
\vskip 1cm
{{\large \bf Fig.3}}: The dependence of the fraction of the hard diffraction
processes induced by instantons from the variable $x=-t/M_x^2$.
\vskip 1cm
{{\large \bf Fig.4}}: The dependence of  the
instantons induced contribution to the DIS structure $F_2^{ep}(x)$ from the
Bjorken variable $x$.
\vskip 1cm
{{\large \bf Fig.5}}: The dependence of the one-spin asymmetry
in the hard diffractive processes from variable $x=-t/M_x^2$. Dashed line is
the result of the calculation without instanton contribution.
\end{document}